\documentclass[twocolumn,amsmath,aps,superscriptaddress,showpacs,floatfix,a4paper]{revtex4}
\usepackage{amsmath}
\usepackage{graphicx}
\usepackage{epsf}
\usepackage{bm}
\usepackage{latexsym}
\RequirePackage{amsopn}
\RequirePackage{amssymb}
\RequirePackage{amsfonts}
\RequirePackage{amsthm}
\usepackage[]{graphicx}
\usepackage[]{amsmath}

\newcommand{\lb }{{\langle}}
\newcommand{\rb}{{\rangle}}

\begin{document}

\title{Coarse-graining dynamics for convection-diffusion of colloids: Taylor dispersion}
\author{Jimaan San\'e}
\affiliation{Rudolf Peierls Centre for Theoretical Physics,
           1 Keble Road, Oxford OX1 3NP, United Kingdom}
\affiliation{Department of Chemistry, Cambridge University,
           Lensfield Road, Cambridge CB2 1EW, United Kingdom}
\author{Johan T.\ Padding}
\affiliation{Computational Biophysics, University of Twente,
           PO Box 217, 7500 AE, Enschede, The Netherlands}
\author{Ard A.\ Louis}
\affiliation{Rudolf Peierls Centre for Theoretical Physics,
           1 Keble Road, Oxford OX1 3NP, United Kingdom}
\date{\today}

\begin{abstract}

By applying a hybrid Molecular dynamics and mesoscopic simulation
technique, we study the classic convection-diffusion problem of Taylor
dispersion for colloidal discs in confined flow. We carefully consider
the time and length-scales of the underlying colloidal
system. These are, by computational necessity, altered in the
coarse-grained simulation method, but as long as this is carefully
managed, the underlying physics can be correctly interpreted.  We find
that when the disc diameter becomes non-negligible compared to the
diameter of the pipe, there are important corrections to the original
Taylor picture.  For example, the colloids can flow more rapidly than
the underlying fluid, and their Taylor dispersion coefficient is
decreased. The long-time tails in the velocity autocorrelation
functions are altered by the Poiseuille flow.  Some of the conclusions
about coarse-graining the dynamics of colloidal suspensions are
relevant for a wider range of complex fluids.


\end{abstract}

\maketitle

\section{Introduction}

Progress in computer simulations of materials arises from two broad
classes of innovation. The first is better algorithms, for example
more efficient ways to calculate equations of motion (for molecular
dynamics) or to sample phase-space (for Monte Carlo). The second comes
from better coarse-grained models of the underlying materials, that is
descriptions that are simpler and more tractable, but nevertheless
retain the fundamental underlying physics that one is interested in
investigating.  Of course progress also arises simply because computers
are continually getting faster, so even with old models and old
methods, new questions can become accessible.

The focus of this paper will be on new
coarse-graining methods, especially for non-equilibrium properties of
colloidal suspensions.  That coarse-graining is needed becomes
immediately obvious when considering the enormous {\em time} and {\em
  length-scale} differences between mesoscopic colloidal and
microscopic solvent molecules. Even a nano-colloid of only 10 nm in
radius will displace on the order of 140,000 water molecules.
Moreover, such a small colloid would still need about $5 \times 10^{-6}
s$ to diffuse over its radius, while a typical collision time of water
molecules is on the order of $10^{-15}s$.  Simulating just a few
nano-colloids in solution is therefore prohibitively expensive with
any explicit model of water.  On the other hand, it is clear from
experiments that many general properties of colloidal suspensions, such
as phase behaviour and basic non-equilibrium behaviour, do not depend
strongly on the radius $R$ of the particles, even though the number of
water molecules per colloid scales as $R^3$.  This physical fact
immediately suggests that coarse-graining methods which ignore the
fine detail of the solvent should be applicable.

A very fruitful coarse-graining strategy has been to employ effective
potentials that integrate out the solvent, and also other degrees of
freedom, to create an intuitively appealing picture of colloids as
``giant atoms''\cite{Loui01a,Liko01}.  Besides rationalising many
results for equilibrium colloids, this viewpoint has stimulated the
use of colloidal systems as a ``playground'' for probing some of the
fundamentals of condensed matter physics.  Colloids have the advantage
that they are much larger than atoms, so that they are much more
easily visualised. At the same time their length-scales are much
longer, so that processes in the time-domain can be more easily
investigated.  Examples of what can be studied include the kinetics of
crystallisation\cite{Auer01}, the dynamics of supercooled
liquids\cite{Week00}, glass formation \cite{Kege00,Pham02} interfacial
phenomena\cite{Aart04}, dislocation motion \cite{Scha04} and surface
melting\cite{Dull04}.   

General processes in atomic and molecular materials where the
non-equilibrium behaviour is dominated by the crossing of potential
barriers, such as nucleation, glass formation and dislocation motion
in crystals, are similar to the concomitant processes in colloidal systems.  But at lower
colloidal densities, it becomes less obvious that this should be the
case.  This is mainly because, in contrast to atoms, the colloids are
in a solvent which induces Brownian motion and long-ranged
hydrodynamic interactions (HI)\cite{Russ89,Dhon96}.  These HI may
decay as slowly as $1/r$ and can qualitatively affect the dynamical
behaviour of a suspension.  The question of how to coarse-grain away
the solvent in such a way as to retain both the Brownian fluctuations
and the long-ranged hydrodynamic interactions is a difficult one.  In
a previous paper \cite{Padd06} (henceforth paper I) we presented a
careful analysis of how to achieve this using the stochastic rotation
dynamics (SRD) algorithm of Malevanets and
Kapral\cite{Male99,Male00}. Note that in the literature, this method
is sometimes also called multiparticle collision dynamics, see the
recent review by Kapral\cite{Kapr08}.  SRD has been applied to a wide
number of different systems, including fluid vesicles in shear
flow\cite{Nogu04}, clay-like colloids\cite{Hech05}, sedimentation of
colloids\cite{Padd04,Wyso08} colloidal rods in shear
flow\cite{Ripo08}, knots in viral DNA\cite{Matt09} and other
examples\cite{Kapr08}.

 In the first section of this paper, we
 briefly review the method, and show how it can be used to bridge
the time and length scales of a colloidal system where HI are
important.  A number of the lessons learned should be valid for other
coarse-grained studies of dynamics.  For example, we argue that many
methods effectively use a ``telescoping down'' procedure to compress
the physical time-scales to a more computationally manageable
hierarchy.  Once the simulations are completed, this hierarchy must be
``telescoped out'' in order to make contact with physical reality.  Of
course this cannot be done without losing some information; there is
no such thing as a free lunch. 

To further illustrate how such a coarse-graining method for dynamics
can be used to investigate the combined effects of Brownian motion and
HI on colloids, we study the diffusive behaviour of colloidal discs
confined to a narrow two dimensional channels with a fluid undergoing
Poiseuille flow.  This convection-diffusion problem was first studied
in a classic paper by G.I. Taylor~\cite{Tay}.  He pointed out that if
a solute diffuses between the stream-lines of a flow field, this adds
an axial diffusive component relative to the average flow field, an effect now called
Taylor dispersion.  In his original work, Taylor derived the
asymptotic form of the dispersion coefficient.  Van
den Broeck~\cite{Vandenbro82} later derived an exact expression for
the dispersion valid at all times. Rather surprisingly Taylor found
that the effective spreading constant, or Taylor diffusion coefficient
as we shall refer to it, was inversely proportional to the molecular
diffusion coefficient $D$ and has the form
\begin{equation}
D^* = \frac{\bar{u}^2L^2}{48D}
\end{equation}
where $L$ denotes the channel radius and $\bar{u}$ the average flow
velocity for a three-dimensional pipe.  Further work in the field has lead to the development of a
more generalised Taylor dispersion theory, as illustrated in the
review by Van den Broeck~\cite{Vandenbro90}. Dispersion theory can
have applications in a multitude of disciplines.  Oscillatory flows,
such as those that arise in estuaries and blood flow have been the
subject of intensive study~\cite{Vandenbro90}. The effect of
interparticle interactions, such as chemical reactions and different
boundary conditions are also of interest~\cite{Vandenbro90}.  Much of this work has focussed on
particles whose radius $R$ is much smaller than the channel radius
$a$.  Here we study in particular what happens to Taylor dispersion and
related problems when the colloidal radius is no longer negligible
compared to the pipe radius, building on earlier work by Brenner and
Edwards\cite{Brened}.  In such situations, we find that the colloid
flows faster than the average of the underlying solvent, mainly because the
colloid can only sample the faster flow lines near the centre of the
flow profile.  At the same time, the Taylor dispersion coefficient is
decreased, because the colloid can sample a smaller range of flow
velocities.  For very narrow pipes more subtle hydrodynamic effects
due to direct interactions with the walls start to become important.

The paper is organised as follows: In section II we discuss some
salient points of our coarse-graining scheme for colloids. In section
III we describe Taylor dispersion for point particles, whereas 
section IV considers the case where the colloid radius is no longer
negligible compared to the pipe width.  Finally, in section IV, we
summarise our main conclusions.

\section{Coarse-graining colloidal dispersions}

\subsection{Effective potentials for colloids}

The detailed interactions between colloidal particles can be very
complex and affected by other (smaller) colloidal particles, polymers, ions and
the properties of the underlying solvent.  However, much progress can
be made by integrating out these degrees of freedom to generate
an effective picture with radial potentials\cite{Loui01a,Liko01}.
The advantage of these are that the well-developed statistical
mechanical machinery developed for atomic and molecular potentials
can now be applied to colloidal suspensions.  Of course colloids can also
be more complicated, for example much recent progress has been made on
``colloidal molecules''\cite{Blaa06}, and colloids can easily  have
anisotropic interactions\cite{Loui02b}. But even for these systems,
the picture of effective potentials is a powerful abstraction\cite{Doye07}.

Nevertheless, it must always be kept in mind that these effective
potentials are not potential energy functions in the Hamiltonian
sense, but rather include within their definition statistical averages
over configurations. Their character is therefore that of a
free-energy.  This has a number of implications such as {\em
transferability problems}, where a potential derived in one context
does not perform well in another context.  A good example would be if
the derived potential depends on the overall colloid density.  Then
using it at a different density could lead to errors.

Another, more subtle, but perhaps equally important issue is that of
{\em representability problems}\cite{Loui02}, where the effective
potential may be used to represent a certain subset of physical quantities, but
not a different subset. 

For example, consider 
a one-component colloidal fluid interacting with a three-body
Hamiltonian of the form:
\begin{equation}\label{eq2.1}
H = K + \sum_{i < j}w^{(2)}(r_{ij})
 + \sum_{i <j <k}w^{(3)}({\bf r}_{ij},{\bf r}_{jk},{\bf r}_{ki}),
\end{equation}
where ${\bf r_i}$ denotes the position of particle $i$ and ${\bf
  r}_{ij} = {\bf r}_i-{\bf r}_j$ and $r_{ij} = |{\bf r}_i-{\bf r}_j|$.
$K$ is the kinetic energy operator, $w^{(2)}(r)$ is an isotropic
pairwise additive potential, and $w^{(3)}({\bf r}_{ij},{\bf
  r}_{jk},{\bf r}_{ki})$ is a triplet or three-body potential.
Three-body potentials are expensive and cumbersome to simulate, and so
one might want to coarse-grain them to a simpler isotropic
pair representation.

Henderson~\cite{Hend74} first showed, using arguments very similar to
those used by Hohenberg and Kohn~\cite{Hohe64} in their famous proof
relating the one-body potential to the one-body density (which laid
the foundation for density functional theory), that ``the pair
potential $v(r)$ which gives rise to a radial distribution function
$g(r)$ is unique up to a constant.''. An extended proof for
orientational correlations can be found in a book by Gray and
Gubbins~\cite{Gray84}, while a more rigourous mathematical discussion
is provided by Chayes, Chayes and Lieb~\cite{Chay84}.  This means that, for
a given state-point, the $g(r)$ generated by a Hamiltonian like that of Eq.~(\ref{eq2.1})
can be reproduced by a unique effective pair potential $v_g(r)$.  
We call approaches that derive $v_g(r)$ the structural route to deriving an effective potential.
The difference with the bare-pair potential $w^{(2)}(r)$ can be written as:
\begin{equation}\label{eq2.2b}
\delta v_g(r) = w^{(2)}(r) - v_g(r).
\end{equation}

Similarly, one could derive an effective potential via an energy route, such that the full internal energy of a system 
 a state point $(N,V,T)$, governed by the Hamiltonian~(\ref{eq2.1}), is reproduced by the simpler two-body formula
\begin{equation}\label{eq2.3}
 U(N,V,T) = \frac12 \rho^2 \int d{\bf r}_1 d{\bf r}_2
 g(r_{12})
 v_{U}(r_{12}).
\end{equation}
The difference with the bare-pair potential $w^{(2)}(r)$ can be written as:
\begin{equation}\label{eq2.2c}
\delta v_U(r) = w^{(2)}(r) - v_U(r).
\end{equation}

It is not hard to show that both $v_g(r)$ and $v_U(r)$ depend on the
state point at which they are derived.  Thus if they are used at a
different state-point, one would expect transferability problems.
What is perhaps more worrying is that one can also show that at a
given state-point, $v_g(r)$ and $v_U(r)$ cannot be the same.  For
example, To lowest order in $\rho$ and $w^{(3)}$, the ratio between the
two corrections is \cite{Casa70,Hoef99,Loui02}:
\begin{equation}\label{eq2.9}
\frac{\delta v_U(r)}{\delta v_g(r)} = \frac13 + {\cal
 O}\left((w^{(3)})^2;\rho^2\right).
\end{equation}
Since $v_g(r)$ is unique, it is therefore impossible to represent the
properties of a system governed by the Hamiltonian of
Eq.~(\ref{eq2.1}) by a single pair potential.  These {\em
representability problems} are widespread in coarse-grained
descriptions of soft-matter systems\cite{Loui02}.  They are also
important in other coarse-grained simulations. For example, one of us
recently studied the structural route to derive radially symmetric
potentials for water\cite{John07}.  There $v_g(r)$ and $v_U(r)$ were
explicitly constructed and look very different, as anticipated by by
Eq.~{\ref{eq2.9}).  At some relevant state-points, using $v_g(r)$ to
calculate the virial pressure resulted in a compressibility factor $Z$
that was almost two orders of magnitude larger than that of the
original multi-site water model used to parameterise $v_g(r)$.
Admittedly, it may not be surprising that an isotropic potential
should perform so poorly when the underlying fluid has complex
orientational correlations.  On the other hand, one could derive a
potential $v_P(r)$, designed to correctly predict the virial pressure
of the underlying water model.  But, due to the uniqueness of
$v_g(r)$, this potential would no longer correctly reproduce the pair
correlations.  It is really a case of ``choose your poison''.

 In practice, things may not always be so dire for many colloidal
suspensions.  First of all, one is not usually trying to reproduce all
properties as accurately as one needs to for a molecular fluid like
water.  Secondly, for many hard-core colloids with short-ranged
interactions, the state-dependence of the interactions is not expected
to be too important.  In this case, transferability and
representability problems, which are often linked\cite{Loui02}, are
not expected to be strong.

For more complex soft-matter systems the story is more subtle, see
e.g. a paper by Murtola {\em et al.}\cite{Murt07} for an excellent
discussion for the case of lipid bilayers.  In general, careful
coarse-graining often leads to potentials that are much softer than
normal atomic or molecular fluids~\cite{Loui00,Loui01a,Liko01,Klap04}.
This can have important advantages when performing equilibrium
calculations, because energy barriers are lowered, making, for example,
Monte Carlo sampling more efficient.  When used in dynamical
simulations, however, the correct application of effective potentials
is more difficult to properly derive.  For example, in the ``polymers
as soft-colloids'' picture\cite{Loui00}, the inter-polymer
interactions are so soft that polymers can easily pass through one
another.  This is clearly not physical, and so schemes that use
effective potentials for polymer dynamics must re-introduce
crossability constraints to prevent coarse-grained polymers from
passing through one another~\cite{Padd02}.  Depletion interactions in
a multi component solution also depend on the relative rates of
depletant diffusion coefficients and particle flow
velocities~\cite{Vlie03,Dzub03}.  Thus care must be taken that the
components integrated out can respond much more rapidly than the
colloids move, so that an effective potential picture still
approximately holds.

In this context it is illuminating to consider the statistical
mechanical origins of dissipative particle dyanmics (DPD), which also
relies on soft interactions\cite{Hoog92,Espa95}. As discussed in paper
I, this method includes two separate innovations.  The
first is to use soft-potentials.  As described above, these can
indeed be very useful for speeding up sampling of equilibrium
behaviour.  For non-equilibrium behaviour however, the application of
these potentials remains somewhat obscure.  One hopes that the
significant lowering of energy barriers leads nonetheless to relative
time-scales that are qualitatively correct.  If this can indeed be
shown to be the case, then the soft potentials are a very useful
computational tool.  In practice however, this is often hard to show
with any certainty, which often limits the reliability or physical
relevance of many DPD simulations.

The second independent innovation in DPD arises from the use of a
thermostat that conserves momentum.  Molecular dynamics simulations
are most naturally performed in the microcanonical ensemble, but most
investigators prefer to use other ensembles such as the canonical
ensemble. To achieve this, they implement a thermostat.  The problem is that many thermostats don't  exactly
conserve momentum, and so the hydrodynamic interactions that
would naturally be present in the microcanonical calculations are not
rendered correctly.  The DPD thermostat solves this problem.  However,
it can also be used with other inter-particle interactions, and
doesn't depend on innovation 1 of DPD, the use of soft potentials.   

For colloids in solution however, using the DPD thermostat does not
generate the correct hydrodynamics without a complete simulation of
the underlying solvent. As pointed out before, this is highly
inefficient, and so a different approach must be used, and that will be
the subject of the next section.

\subsection{Coarse-graining dynamics of colloids}

To describe the dynamics of colloidal suspensions, we follow paper I
and implement the SRD method first derived by Malevanets and
Kapral~\cite{Male99}.
 The method works by exploiting the
fact that Navier Stokes hydrodynamics arises from local momentum
conservation.  Thus one can employ greatly simplified dynamics that are
computationally very efficient to simulate, and still generate hydrodynamic behaviour.
There are many methods in this class, including direct simulation
Monte Carlo (DSMC) method of Bird~~\cite{Bird70,Bird94} and the
Lattice Boltzmann (LB) technique where a linearized and pre-averaged
Boltzmann equation is discretised and solved on a
lattice~~\cite{Succ01}. In particular Ladd and others have extended
this method to model colloidal
suspensions~~\cite{Ladd93,Ladd01,Cate04,Loba04,Capu04,Chat05}.  SRD
has the particular advantage that transport coefficients have been
analytically calculated~\cite{Ihle03,Kiku03,Pool04}, greatly
facilitating its use.  It is important to remember that for all these
particle based methods, the particles should not be viewed as some
kind of composite supramolecular fluid units, but rather as
coarse-grained Navier Stokes solvers (with noise in the case of
SRD)~\cite{Padd06}.

The SRD fluid is modelled by $N$ point particles of mass $m$, with
positions ${\bf{r}}_{i}$ and velocities ${\bf{v}}_{i}$. The coarse
graining procedure consists of two steps, streaming and collision. During
the streaming step, the positions of the fluid particles are updated via
\begin{equation}
{\bf{r}}_{i}(t+\delta t_{c}) = {\bf{r}}_{i}(t) + {\bf{v}}_{i}(t)\delta t_{c}.
\end{equation} In the collision step, the particles are split up into
cells with sides of length $a_0$, and their velocities are rotated
around an angle $\alpha$ with respect to the cell centre
of mass velocity,
\begin{equation}
{\bf{v}}_{i}(t+\delta t_{c}) = {\bf{v}}_{c.m,i}(t) +
\mathcal{R}_i(\alpha)\left[{\bf{v}}_{i}(t)-{\bf{v}}_{c.m,i}(t)\right]
\end{equation}
where ${\bf{v}}_{c.m,i}=\sum^{i,t}_{j}(m{\bf{v}}_{j})/\sum_{j}m$ is
the centre of mass velocity of the particles the cell to which $i$
belongs, $\mathcal{R}_i(\alpha)$ is the cell rotational matrix and
$\delta t_{c}$ is the interval between collisions. The purpose of this
collision step is to transfer momentum between the fluid particles
while conserving the energy and momentum of each cell.  The algorithm
is efficient because direct interactions between the particles are not
taken into account.

Spherical colloids of mass $M$ can be embedded in an SRD solvent using
a Molecular Dynamics technique, as first shown by Malevanets and
Kapral~\cite{Male00}, and studied in detail in reference I.
We employ the following scheme:
For the colloid-colloid interaction we
use a standard steeply repulsive potential of the form:
\begin{displaymath}
\varphi_{cc}(r) = \left\{
\begin{array}{ll}
4\epsilon\left( \left(\frac{\sigma_{cc}}{r}\right)^{48}-
\left(\frac{\sigma_{cc}}{r}\right)^{12} +\frac{1}{4}\right) & (r\leq 2^{1/24}\sigma_{cc})\\
0 & (r\geq 2^{1/24}\sigma_{cc})
\end{array} \right.
\end{displaymath}
while the interaction between the colloid and the solvent is described by
a similar, but less steep, potential:
\begin{displaymath}
\varphi_{cs}(r) = \left\{
\begin{array}{ll}
4\epsilon\left( \left(\frac{\sigma_{cs}}{r}\right)^{12}-
\left(\frac{\sigma_{cs}}{r}\right)^{6} +\frac{1}{4}\right) & (r\leq 2^{1/6}\sigma_{cs})\\
0 & (r\geq 2^{1/6}\sigma_{cs})
\end{array} \right.
\end{displaymath}
where $\sigma_{cc}$ and $\sigma_{cs}$ are the colloid-colloid and
colloid-solvent collision diameters.  We propagate the ensuing equations of
motion with a Velocity Verlet algorithm using a molecular
dynamic time step $\Delta t$
\begin{eqnarray} R_{i}(t+\Delta t) &=&
  R_{i}(t) + V_{i}(t)\Delta t + \frac{F_{i}(t)}{2M}\Delta t^{2}\\
  V_{i}(t+\Delta t) &=& V_{i}(t) + \frac{F_{i}(t)+F(t+\Delta
    t)}{2M}\Delta t
\end{eqnarray}
where $R_{i}$ and $V_{i}$ are the position and velocity of the
colloid, and $F_{i}$ the total force exerted on the colloid.  Coupling
the colloids in this way leads to slip boundary conditions.  Stick
boundary conditions can also be implemented~\cite{Padd05}, but for
qualitative behaviour, we don't expect there to be important
differences.  In parallel the velocities and positions of the SRD
particles are streamed in the external potential given by the colloids
and the external walls and updated with the SRD rotation-collision
step every time-step $\delta t_c$.  We chose $\sigma_{cc} = 4.3
a_0$. The choice of time-steps, as well as a number of other technical
issues are described in more detail in ref I, and for the case of
two-dimensional discs, in\cite{Sane08}.

One key step in the implementation and interpretation of these kinds
of coarse-grained dynamical simulations lies in the realisation that
hydrodynamic phenomena depend primarily on a set of dimensionless
numbers, and that it is important to keep these in the right regimes
in order to simulate the correct physical behaviour.  Typical numbers
include the Reynolds number Re which measures the relative strength of
viscous and inertial forces, and the Mach number Ma, defined as the ratio of a
typical colloidal velocity to the velocity of sound $c_s$ of the underlying
fluid.  In a real colloidal system, the Re and Ma numbers are very
small, typically on the order of $10^{-6}$ or less.  However, as long
as one keeps them less than about $10^{-1}$, the physical regime
doesn't really change much, a fact that can be exploited in
simulations.  For example if the real Ma number of a colloidal
suspension were reproduced, it would mean that millions of sound waves
would need to be resolved for modest motion of the colloids.  But to
be in the right physical regime, a sound velocity well separated from
other velocities is all that is needed.  Obviously having a slower
sound speed greatly increases the efficiency of the simulation.  The
example of the Ma number can be extended to a whole range of other
hydrodynamic numbers, as discussed in more detail in reference
I\cite{Padd06}.

A related set of arguments can be made for the time-scales of a
typical colloidal system.  A colloid of radius $1 \mu m$ diffuses over
its radius in a time of order several seconds.  At the same time, the
shortest physically relevant time-scale for the colloidal dynamics is
the Fokker-Planck time $\tau_{FP}$, defined in \cite{Dhon96} as the
time over which the colloidal force-force correlation function decays.
For a typical colloid, this may be as small as $10^{-13}s$. Other
important time-scales include the sonic time-scale $t_{cs}=R/c_s$, on
the order of $10^{-10} s$ for a typical colloid. The kinematic
time-scale $\tau_\nu = R^2/\nu$ which measures the time for vorticity,
which diffuses with the kinematic viscosity $\nu$, to diffuse over one
colloidal radius is also important, and linked to the Re number.  In a
real colloidal system, these time-scales are all separated from each
other by many orders of magnitude.  However, to be in the correct
physical regime, one doesn't necessarily need to resolve all those
time-scales.  All that is really needed is a clear time-scale
separation.

 Our strategy is then as follows: First, the relevant time-scales are
 identified.  These are then {\em telescoped} down to a hierarchy
 which is compacted to maximise simulation efficiency, but
 sufficiently separated to correctly resolve the underlying physical
 behaviour.  We illustrate this process  in
 Figure~\ref{fig:telescope}, and discuss it in more detail in ref
 I.  Although the particular example studied here
 concerns a colloid in suspension, we argue that many other
 coarse-graining methods make implicit use of some aspect of
 telescoping down.  A careful time-scale analysis is very important
 for the correct physical interpretation of the simulation results.   

\begin{figure}[t]
  \scalebox{0.4}{\includegraphics{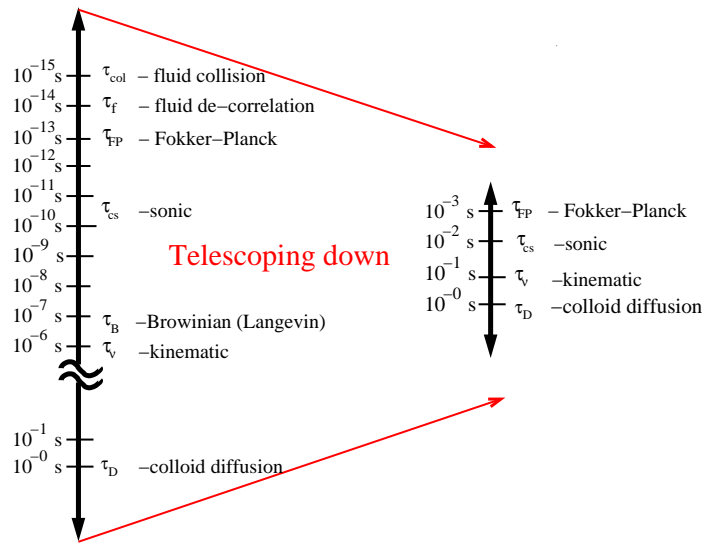}}
\caption{ Telescoping down: The hierarchy of time-scales for a
  colloid (here the example taken is for a colloid of radius 1 $\mu m$
  in H$_2$0) is compressed in the coarse-grained simulations to a more
  manageable separation.  As long as the physically important times are clearly separated, the
  simulation should still generate the correct physical picture.  Once
  the simulations are completed, they can be related in more detail to
  particular experiments by telescoping back out to the relevant
  experimental time-scales.
\label{fig:telescope}
}
\end{figure}

Having sketched our simulation technique, we now turn to an
application where  diffusion, convection, and hydrodynamic
interactions all play a role.

\section{Taylor dispersion for small solutes}

When a fluid flows between two parallel plates, stick boundary conditions mean that the local fluid velocity is zero at the plate surfaces. 
The resulting  Poiseuille flow has a velocity distribution of the form~\cite{Happ73}:
\begin{equation}
u(y) = \frac{3}{2}\bar{u}\left(1-\frac{y^{2}}{L^{2}}\right),
\end{equation}
where $\bar{u}$ is the mean velocity, $2L$ is the plate separation, and
$y$ is a coordinate measuring the distance from the centre of the
pipe.  Thus the maximum velocity $u_0=\frac32 \bar{u}$ is at the centre of the flow profile.

The basic idea behind Taylor dispersion is that a solute particle will
diffuse between the streamlines.  Its average position moves forward
with the average flow velocity, but because some streamlines move faster than the average, and some slower, the diffusion
between streamlines adds an extra dispersive component relative to its average motion in the axial direction.

\begin{figure}
\begin{center}
\includegraphics[width=0.5\textwidth]{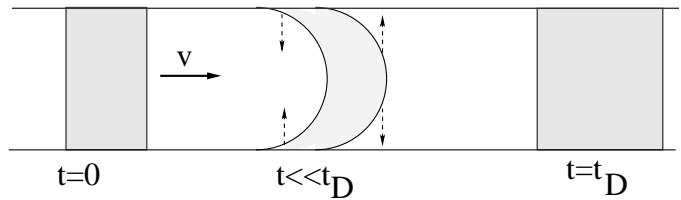}
\caption{\label{Fig:Poiseuille}Taylor dispersion between parallel plates.
The initial solute gets stretched to a paraboloid
shaped plug. Diffusion, indicated by the vertical arrows, evens out the
concentration profile leading to a wider plug.  Taylor dispersion describes the diffusive spreading of the plug.} 
\end{center}
\end{figure}

To get a better physical understanding of Taylor diffusion, we follow
an argument from~\cite{Squires}, depicted in fig.\ref{Fig:Poiseuille}.
For time scales less than $t_{D}\sim L^{2}/D$, convection initially
stretches the solute into a parabola. The solute at the front end of
the parabola leads the solute at the edges by a distance $u_{0}t$
after a time $t$.  Diffusion across the channel then smears the
parabola into a plug in a time-scale $t_{D}$ it takes the
solute to reach the edges. Thus the parabola is smeared into a plug of
width $W_{tD}\sim u_{0}t_{D}=u_{0}L^{2}/D$.  We can describe this as
the the plug taking a random step of size $\sim W_{tD}$ at each `step'
$t_{D}$. If we now repeat the process for different stripes of the
plug; each convectively stretched then diffusively smeared, then after
$N$ steps, the plug will have evolved as
\begin{equation}
\langle W^{2} \rangle \sim N W_{tD} \sim
\left(\frac{u_{0}^{2}L^{2}}{D}t\right).
\end{equation}
Accordingly, in addition to molecular diffusivity $D$, the solute can be seen
to diffuse along the channel, with respect to the average flow, with an
effective dispersion coefficient
\begin{equation}
D^{*} \sim \frac{u_{0}^{2}L^{2}}{D}.
\end{equation}

Note that the effective dispersion coefficient is inversely proportional to the
molecular diffusion coefficient, which may seem counter intuitive at first
glance. However, solute molecules with higher molecular diffusivities spend
more time diffusing across the pipe and less time sampling particular velocity
streamlines thus reducing their \emph{effective} spreading.

A more sophisticated calculation based on flow in a three dimensional pipe of radius $L$ yields a prefactor~\cite{Probs}
\begin{equation}
D^{*} = \frac{\bar{u}^{2}L^{2}}{48D}.
\end{equation}
In previous work \cite{PhD} we have calculated the exact value for the
prefactor in two dimensions. By integrating the equations of motion over the
plate separation and implementing the appropriate boundary conditions we found 
that
\begin{equation}\label{phd1}
D^{*} = \frac{2\bar{u}^{2}L^{2}}{105D}.
\end{equation}
In this paper we will carry out most of our simulations and analysis
in two dimensions, simply because simulations are easier, and the
fundamental physics we are after should not be very dependent on the
exact dimension we work in.

The calculations above assume that the molecular diffusion in the
axial direction is negligible.  When this is not the case, a good
approximation for the Taylor dispersion coefficient was first given by Aris~\cite{Aris} to be:
\begin{equation}
D_{TA} = D^{*} + D,
\end{equation} and so
$D_{TA}$ is often called the Taylor-Aris dispersion coefficient.

Taylor dispersion has been extensively used to study the diffusion
coefficients of small molecular solutes\cite{Cast94} where $D$ is
typically $10^{-8}cm^{2}/s$ so that $D^*$ is expected to be large, making
it easy to measure. Experimental setups examining atomic Taylor
dispersion use tubes of order a few $mm$ in diameter and lengths many meters.  In such setups, $D$ can be measured with an
accuracy of $\pm1\%$~\cite{Vandenbro90}.  For colloidal systems, where  diffusion coefficients are much smaller, it may be possible to use a
shorter tube, because the Taylor-Aris dispersion coefficient would in
fact be much larger.

\begin{figure}\label{brenner3}
\includegraphics[width=0.5\textwidth]{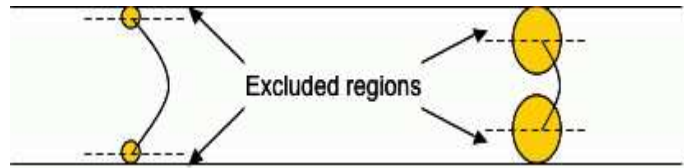}
\caption{Finite size effect correction to the flow profile. Colloids are no
longer able to velocity gradients near the walls resulting
in an average velocity higher than that of the solvent.}
\end{figure}

\section{Taylor dispersion of finite sized colloidal discs}

As illustrated in Fig.~\ref{brenner3}, when the colloid radius becomes
comparable to $L$, the colloids can no longer sample the entire
Poiseuille flow profile.  This leads to a number of
corrections~\cite{Brened}.  Firstly, because the colloids are excluded
from the slower flow-lines, their average velocity is faster than the
underlying flow of the fluid.  This can easily be calculated for
Poiseuille flow
\begin{equation}\label{eq:vel}
u_{c} =
\frac{3}{2}\frac{\bar{u}}{L-R}\int^{L-R}_{0}
\left(1-\frac{y^{2}}{L^{2}}\right)dy=
\frac{3}{2}\bar{u}\left(\frac{2}{3}+
\frac{2}{3}\chi-\frac{1}{3}\chi^{2}\right).
\end{equation}
where $\chi = R/L$ measures the size of the colloid compared to the
distance between the plates.  For infinitely small particles with
$\chi=0$ the average velocity is $u_c=\bar{u}$, but as $\chi$ increases,
so does the average velocity, until it reaches a maximum of $u_c =
\frac32 \bar{u}$ at $\chi=1$.  This size exclusion phenomenon can used
to analyse the relative diameters of colloidal particles (hydrodynamic
chromatography~\cite{Mchugh}).

For finite $\chi$, the Taylor dispersion coefficient also undergoes
corrections\cite{Brened}.  If the new boundary conditions are taken
into account, then our calculation of Eq.(\ref{phd1}) incurs further
corrections~\cite{PhD}:
\begin{equation}\label{eq:Brenner} D^{*} =
\frac{\bar{u}^{2}L^{2}}{D}\left(\frac{(1-\chi)^{4}}{30} -
\frac{(1-\chi)^{6}}{70}\right)
\end{equation}
for $2d$ flows.  The radial dispersion of the colloids is now
effectively reduced as they are unable to sample the higher velocity
gradients near the edges, reducing the radial dispersivity and leading
to a lower value of the Taylor dispersion coefficient.

Note, however, that the calculations above assumed that {\bf 1}, the
colloids perfectly follow the stream-lines, and are not themselves
affected by their proximity to the plates, and {\bf 2}, that the
average flow profile is not itself perturbed by the presence of the
colloids.  For three dimensions, Brenner $\&$ Gaydos~\cite{Brengay}
used a moment analysis to show that extra terms enter into
Eq.(\ref{eq:vel}) that reduce the numerical coefficients so that
the actual flow velocity is lower than a calculation which ignores the
wall effects.  What happens to the average flow in two dimensions, and
how this exactly affects the Taylor dispersion beyond that shown in
Eq.~(\ref{eq:Brenner}), is not known.

\begin{figure}\label{taymov}
\includegraphics[width=0.5\textwidth]{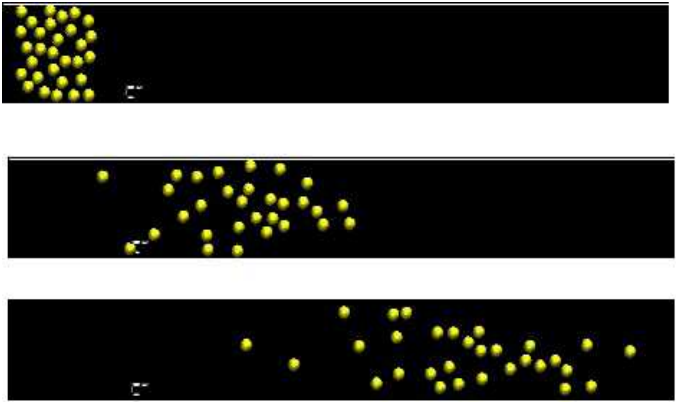}
\caption{Simulation snapshots at different time-interals of the Taylor diffusion of colloids in a flow field.}
\end{figure}

To study the effect of finite particle size we performed computer
simulations of colloids and heavier SRD tracer particles using the
methodology described in papers I and~\cite{Sane08}.
Fig.~\ref{taymov} shows some typical snapshots of colloidal dispersion
driven by Taylor dispersion.  We take care to make sure that the Re
number is always less than about $0.1$, and that other hydrodynamic
numbers are also in the correct regime.  We simulate with up to $60$
colloids and for box lengths of $238 R$.  Averages are measured after the
transients are deemed to have decayed away.

\begin{figure}
\begin{center}
\includegraphics[width=0.5\textwidth,height=0.5\textwidth,angle=-90]{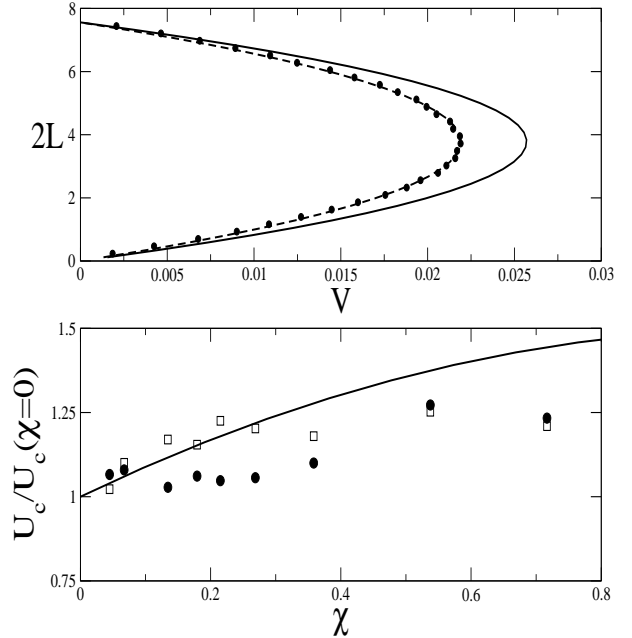}
\caption{\label{Fig:fitviscwall} Top : Effect of the colloid
  concentration on the solvent flow profile. The straight line
  represents the theoretical flow profile for a pure SRD solvent.  The
  simulation points represent the flow profile for a homogeneous
  distribution of colloids at a packing fraction of $\phi=0.1$ while
  the dotted line shows the fit with the increased viscosity. Bottom
  : The average flow velocities of colloids (circles) and ``ideal''
  colloids (squares) are compared to the original simple
  prediction of Eq.~\protect\ref{eq:vel} that ignores the effects of
  the walls.  As expected, for increasing $\chi = R/L$,  the walls on average slow down the
  particles compared to what one would expect if they just followed
  the unperturbed Poiseuille flow lines. }
\end{center}
\end{figure}

One effect that the colloids can have on the flow is that they
increase the viscosity, which in turn affects the average velocity.
This is shown in the top panel of Fig.~\ref{Fig:fitviscwall}. By
fitting the measured profile for different colloid packing fractions
$\phi$ we find that the viscosity of the mixture scales roughly as
$\eta_{\phi} \sim (1+1.95\phi)\eta_{0}$.  In three dimensions and in
the bulk, the pre-factor in front of $\phi$ is $2.5$~\cite{Russ89,Dhon96}.

In the bottom panel of Fig.~\ref{Fig:fitviscwall} we show the average
velocity of colloidal particles as a function of the radius to
pipe-size ratio $\chi$.  These show a decrease of their velocity
compared to the ideal calculation of Eq.~(\ref{eq:vel}) due to the
effect of hydrodynamic interactions with the walls.  We also tested
the average velocity of SRD tracer particles (``ideal colloids''). These have a larger mass
than the other fluid point particles, and in addition, interact with
the walls with an interaction of radius $R$.  Othewise they are
point-particles and participate in the SRD collision rules. We expect
that these particles have a smaller effect on the fluid flow profile
than the colloids do.  As shown in Fig.~\ref{Fig:fitviscwall}, these
particles also don't follow the ideal correction term, and so
presumably also experience hydrodynamic interactions with the wall
that slow them down.

\begin{figure}
\begin{center}
\includegraphics[width=0.5\textwidth,height=0.5\textwidth,angle=-90]{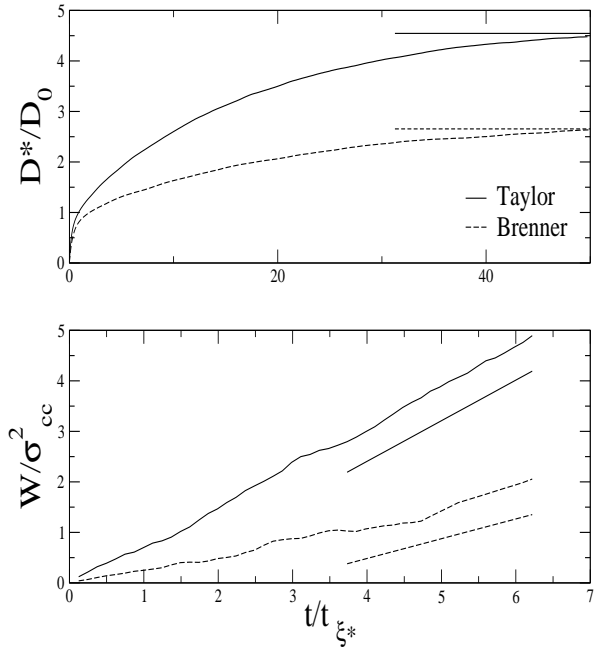}
\caption{\label{Fig:taydepl} Top panel: Integral of the velocity fluctuations.
Bottom panel: Effective mean square displacement with respect to the average
flow. Both graphs are for ``ideal colloids'' made up of heavier SRD particles.
}
\end{center}
\end{figure}

Within a simple Langevin picture, a colloid loses memory its velocity
on a time-scale given by $t_\xi = M/\xi$, where the friction
coefficient $\xi$ is defined by $D=k_B T/\xi$.  Although for colloids
this Langevin picture is in fact heavily misleading, see e.g.\ the
discussion in the appendix of ref I\cite{Padd06}, for our purposes
here it should suffice as a useful reference time-scale.

If the fluctuation-dissipation theorem is invoked, then an effective
Taylor dispersion friction $\xi^{*}$ can be defined, such that
$D^{*}=\frac{k_{B}T}{\xi^{*}}$.  In a simple Langevin picture,
fluctuations will die out on the time scale
\begin{equation}\label{xi}
t_{\xi^{*}} = \frac{M}{\xi^{*}} = \frac{MD^{*}}{k_{B}T} =
\frac{2}{105}\frac{MD}{k_{B}T}\frac{\bar{u}^{2}L^{2}}{D^{2}} =
\frac{2}{105}t_{\xi}Pe^{2}
\end{equation}
where we have defined the pipe Peclet number
\begin{equation}\label{Pe}
Pe = \frac{\bar u L}{D} = t_U/\tau_{DL}
\end{equation}
which can be viewed as the ratio of the time $t_U=L/\bar{u}$ it takes
the fluid to move in the axial direction over a distance L, to the
time $\tau_{DL}$ it takes a particle to diffuse that distance.

To measure Taylor dispersion as accurately as possible, we simulated
`ideal' colloids of mass $M=62.3m$ under flow and measured the
dispersion of 25 of these in a pipe of width such that $\chi = 0.134$.
A separate set of simulations was done in the same pipe, but with
``ideal'' SRD colloids with the same wall-colloid interaction as the
solvent.  For these ideal particles and the relatively wide pipe, we
don't expect strong wall effects, so that the simple
Eq.~(\ref{eq:Brenner}) should provide a good approximation to the
Taylor dispersion.  The results of these simulations are shown in
Fig.\ref{Fig:taydepl}.  The plain curves denote results for the
heavier SRD fluid particles without an extra wall-interaction, while
the dashed curves depict results for ``ideal colloids'', i.e.\ the
heavier SRD particles with an extra interaction with the wall. In the
top section of the graph, we have plotted the integral of the velocity
auto-correlation function for both sets of particles.  The horizontal
lines are a guide to the eye and intersect the $y$-abscissas at the
theoretical values for dispersion predicted in Eq.~(\ref{eq:Brenner})
for $\chi =0$ and $\chi=0.134$ respectively.  In both instances, we
see that the simulated integrals plateau near the lines suggesting
good agreement with Taylor theory, including the corrections of
Eq.~(\ref{eq:Brenner}).  The lower part of Fig.\ref{Fig:taydepl} shows
the mean squared displacement of the particles w.r.t.\ the average of
the centre of mass of a large number of colloids.  This is expected to
scale as $X^2 \sim 2 D^* t$.  Note that even for this relatively small
value of $\chi$, the Taylor dispersion coefficient is down by about
$\frac23$, so that the effect of a finite size is much stronger on the
Taylor dispersion than it is on the relative velocity of the
particles, which only shows a very modest correction for that value of
$\chi$.

Fig.\ref{Fig:msdTaylor} depicts the mean square displacement w.r.t.\ the centre of mass flow of normal colloidal discs.  The measured Taylor
dispersion coefficient $D^*$ appears to lie between the predictions of
Eq.~(\ref{phd1}) and (\ref{eq:Brenner}).

\begin{figure}
\begin{center}
\includegraphics[angle=-90, width=0.5\textwidth]{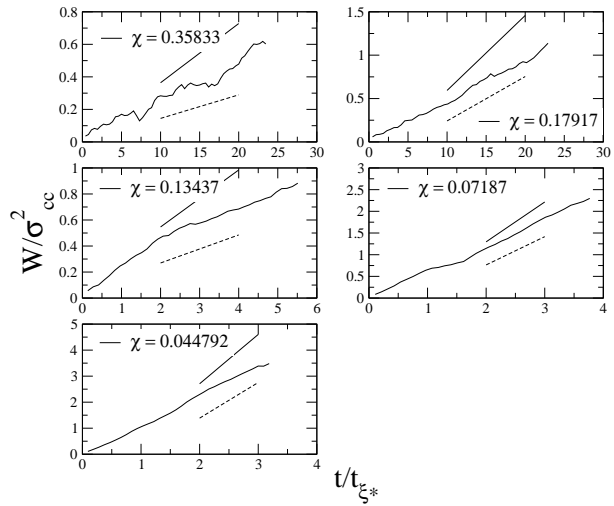}
\caption{\label{Fig:msdTaylor} Effective mean square displacement of
  colloids undergoing Taylor dispersion. Simulations were carried out
  in pipes of sizes $\chi = 0.358,0.179,0.134,0.0719,0.0448$
  respectively.  The solid straight line represents the approximation
  for Taylor dispersion from Eq.~(\protect\ref{phd1}), and the dotted
  lines denote the finite size correction predicted in
  Eq.~(\protect\ref{eq:Brenner}).  Simulation parameters were chosen
  to ensure that the self and Taylor dispersion are always clearly
  separated, i.e. that $D^{*}/D = t_{\xi^{*}}/t_{\xi} \geq 5$. All
  runs were carried out for colloid Reynolds number of less than
  $0.1$.}
\end{center}
\end{figure}

\begin{figure}
\includegraphics[width=0.5\textwidth,height=0.5\textwidth,angle=-90]{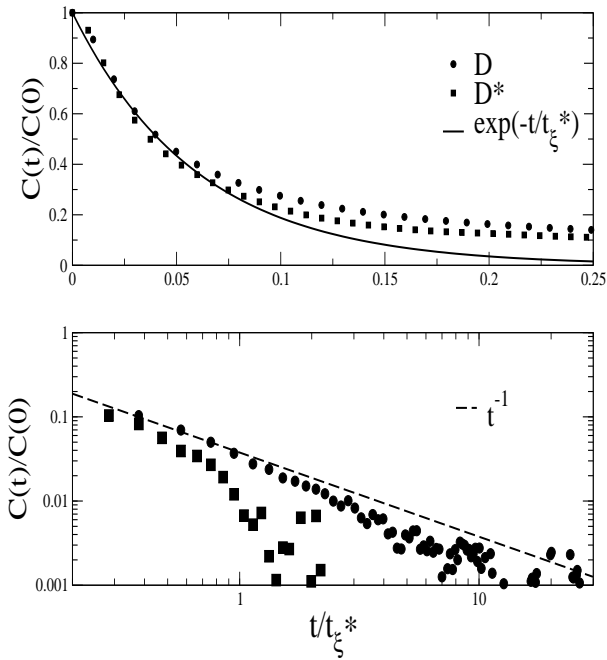}
\caption{\label{Fig:vacttay}Decay of the velocity autocorrelation
  function for particles undergoing normal (circles) and Taylor
  (squares) dispersion.  The solid line in the top panel depicts the
  Enskog exponential decay. The dotted line in the bottom panel shows
  $t^{-1}$ decay and serves as a guide to the eye.  }
\end{figure}

We also investigated the behaviour of the velocity autocorrelation
function(VACF) $C(t)=\lb\delta v(t)\delta v(0)\rb$ for colloids
undergoing Taylor dispersion in a pipe of width $\chi = 0.13437$.
These are shown in Fig.\ref{Fig:vacttay}, and compared to the VACF for
particles for the case of no flow.  For short times both curves are
well described by a simple Enskog kinetic equation (see
ref~\cite{Sane08}).  Physically this is to be expected, as these
kinetic processes occur on the sonic time-scale which is much faster
than the flow, and so it is gratifying to see that our simulations
correctly separate out this time-scale from the others.  At long times
the two VACFs differ.  As shown in the bottom panel, the particle
without flow shows the expected $t^{-1}$ long time tail, first
observed by Alder and Wainwright\cite{Alde70}.  This tail is caused by
hydrodynamic correlations that propagate through the fluid.  For a
finite sized pipe, we expect the tail to feel the influence of the
walls at longer times\cite{Pago01,Sane08}, and some evidence of this
may be visible. Nevertheless, this tail clearly differs from that of
the VACF for Taylor dispersion, which shows a much more rapid
non-algebraic decay.  Instead it decays on a time scale comparable to
$t_{\xi*}$. The reason for this is most likely that the hydrodynamic
correlations are washed out by the effect of the flow.

\section{Summary}

In summary then, we have adapted a hybrid MD-SRD technique to study
colloidal discs in two dimensions undergoing flow in a narrow pipe.
By carefully considering the time and length-scales involved in the
problem, we were able to derive a coarse-graining method that is
tractable, but still resolves the main physical properties under
investigation. We measured corrections to the average flow profile,
and find that the colloids move faster than the flow when the ratio
$\chi = R/L$ increases, but that hydrodynamic wall effects also slow
down the colloids when the pipes become very narrow.  Finite $\chi$
has an even bigger effect on the Taylor dispersion coefficient,
decreasing its value from that expected for $\chi=0$.  We also
measured the VACF under flow conditions, and find that this decays
more rapidly than the case of no flow, most likely because the Taylor
dispersion process breaks up the long-time hydrodynamic correlations.

Finally, given that we used a ``telescoping down'' of the hierarchy of
time-scales to derive a tractable simulation scheme for the dynamics
of our colloidal system, how would one ``telescope up'' to make
experimentally relevant predictions?  The first step is to cast
properties of interest into dimensionless form as much as possible.
From that we see, for example, that the relative decrease of Taylor
dispersion depends only on the ratio $\chi$, and not on the absolute
value of either the pipe size or the colloid radius $R$.  Thus many
different experimental colloid and pipe sizes should be describable by
the same set of simulations.  On the other hand, if we are interested
in the long-time tails, then a more subtle analysis is needed.  It
appears that for colloids undergoing flow, the VACF decays on a time
of the order of $t_{\xi*}*$, which, through Eq.~(\ref{xi}), can be
related to the Langevin time $t_\xi$ and the $Pe$ number.  Again, one
can rather straightforwardly estimate $t_\xi$ for colloids if one
knows their diffusion coefficient.  However, the relative values of
$t_\xi$ and the diffusion time $\tau_D$ that a colloid needs to diffuse
over its own radius may be very different in the simulation than in an
experiment. That means that the two time-scales in the simulation need
to each be scaled differently when comparing to experiment.  So in the
end, to interpret a coarse-grained simulation, the best advice is to
try and understand as best as possible the underlying physics.

\acknowledgements J.S. thanks Schlumberger Cambridge Research and
IMPACT FARADAY for an EPSRC CASE studentship which supported this
work. A.A.L. thanks the Royal Society (London) and J.T.P. thanks the
Netherlands Organization for Scientific Research (NWO) for financial
support.  We thank H. Lekkerkerker for first suggesting to us that we
study Taylor dispersion and R. Castillo for helpful discussions about
Taylor dispersion experiments.

\end{document}